\documentclass[preprint,aps,eqsecnum,graphicx]{revtex}
\def\red{
}
\def\black{
}

\def\mycomm#1{\hfill\break\strut\kern-3em{\red\tt ====> #1\black}\hfill\break}

\def\ds{\displaystyle}

\def\meddeepstrut{\vrule height 1.5ex depth 1.0ex width 1pt}

\def\tallstrut{\vrule height 4.5ex depth 1.5ex width 0pt}

\def\verytallstrut{\vrule height 6.5ex depth 1.5ex width 0pt}
\def\eqref#1{(\ref{#1})}

\def\MUU{ {\cal V } }
\def\MUD{ {\cal P } }

\makeatletter
\def\hlinewd#1{\noalign{\ifnum0=`}\fi
\hrule \@height #1 \futurelet \reserved@a\@xhline}
\def\hwhiteline{\noalign
{\ifnum0=`}\fi\hrule
\@height 0pt\vskip 1.0ex\futurelet \reserved@a\@xhline}
\makeatother
\def\gray{\special{ps: 0.40 setgray}}
\def\black{\special{ps: 0.0 setgray}}

\newcommand{\mydraft}{
\newcount\timecount
\newcount\hours \newcount\minutes  \newcount\temp \newcount\pmhours

\hours = \time
\divide\hours by 60
\temp = \hours
\multiply\temp by 60
\minutes = \time
\advance\minutes by -\temp
\def\hour{\the\hours}
\def\minute{\ifnum\minutes<10 0\the\minutes
    \else\the\minutes\fi}
\def\clock{
\ifnum\hours=0 12:\minute\ AM
\else\ifnum\hours<12 \hour:\minute\ AM
\else\ifnum\hours=12 12:\minute\ PM
    \else\ifnum\hours>12
     \pmhours=\hours
     \advance\pmhours by -12
     \the\pmhours:\minute\ PM
     \fi
    \fi
\fi
\fi
}
\def\fullclock{\hour:\minute}
\begin{centering}
\gray
\font\Hugett  =cmtt12 scaled\magstep4
\hbox{\Hugett Draft:\today,\clock}
\black
\end{centering}
\vskip -1.7cm
$\phantom{a}$
} % end of \draft definition
\def\beq#1{\begin{equation} \label{#1}}
\def\eeq{\end{equation}}

\def\bea{\begin{eqnarray}}
\def\eea{\end{eqnarray}}

\newskip\humongous \humongous=0pt plus 1000pt minus 1000pt

\newif\ifdtup

\begin{document}

{\tighten
\preprint {\vbox{
\hbox{$\phantom{aaa}$}
\vskip-0.5cm
%\hbox{\today}
%\hbox{}
%
\hbox{TAUP 2841/06}
\hbox{WIS/17/06-NOV-DPP}
\hbox{ANL-HEP-PR--06-86}
\hbox{$\phantom{a}$}
}}

\title{The New $\Sigma_b$ multiplet, 
heavy baryon mass predictions,
\\
meson-baryon universality 
\\
and
\\
 effective supersymmetry
in hadron spectrum
\\
} 

\author{Marek Karliner\,$^{a}$\thanks{e-mail: \tt marek@proton.tau.ac.il}
\\
and
\\
Harry J. Lipkin\,$^{a,b}$\thanks{e-mail: \tt
ftlipkin@weizmann.ac.il} }
\address{ \vbox{\vskip 0.truecm}
% $^a\;$Cavendish Laboratory\\
% Cambridge University, England;\\
%\mbox{}\\
% and\\
$^a\;$School of Physics and Astronomy \\
Raymond and Beverly Sackler Faculty of Exact Sciences \\
Tel Aviv University, Tel Aviv, Israel\\
\vbox{\vskip 0.0truecm}
$^b\;$Department of Particle Physics \\
Weizmann Institute of Science, Rehovot 76100, Israel \\
and\\
Physics Division, Argonne National Laboratory \\
Argonne, IL 60439-4815, USA\\
}
\maketitle
\vskip1.0cm

%\mydraft
\begin{abstract}%
The recent measurement by CDF
\ $M(\Sigma_b) - M(\Lambda_b) = 192 \,{\rm MeV}$ \
is in striking agreement 
with our theoretical prediction
$M(\Sigma_b) - M(\Lambda_b) = 194 \,{\rm MeV}$.
In addition,
the measured splitting
$M(\Sigma_b^*)- M(\Sigma_b) = 21$ MeV agrees well with the predicted splitting
of 22 MeV.
We point out the connection between these predictions and an
effective supersymmetry between
mesons and baryons related by replacing a light antiquark by a light
diquark. 
We discuss the theoretical framework behind these predictions
and use it to provide additional predictions for the masses of
spin-${1\over2}$ and spin-${3\over2}$ baryons containing heavy quarks,
as well as for magnetic moments of 
$\Lambda_b$ and $\Lambda_c$.
\end{abstract}
% end tighten

\vfill\eject

% label footnotes by symbols, rather than numbers
\renewcommand{\thefootnote}{\fnsymbol{footnote}}
\setcounter{footnote}{1}

\section{Striking Agreement with Meson-Baryon Predictions}

A new challenge demanding explanation from QCD is posed by 
the remarkable agreement shown in Fig.~1 between the experimental masses 
5808 MeV and 5816 MeV of the newly discovered  $\Sigma_b^+$ and $\Sigma_b^-$ 
and the 5814 MeV quark model 
prediction\cite{NewPenta} from meson masses 
\hfill\break
 \beq{newpred}
{{M_{\Sigma_b} - M_{\Lambda_b}}\over{M_{\Sigma} - M_\Lambda}} =  
{{(M_\rho - M_\pi)-(M_{B^*}-M_B)}\over{(M_\rho - M_\pi)-(M_{K^*}-M_K)}}= 2.51
\end{equation}
\hfill\break
\noindent
This then predicts that 
the isospin-averaged
mass splitting is
$M_{\Sigma_b} - M_{\Lambda_b} = 194 \,{\rm MeV}$
and
$M(\Sigma_b) = 5814$ MeV,
using
the most recent CDF $\Lambda_b$ mass measurement
\cite{CDF_Lambda_b}
$M(\Lambda_b) = 5619.7 \pm 1.2\,\hbox{(stat.)}\pm1.2\,
\hbox{(syst.)\ MeV}$.\footnote{Ref.~\cite{NewPenta}
used an older value $M(\Lambda_b) = 5624$ MeV \cite{RPP}, yielding
$M(\Sigma_b) = 5818$ MeV.}

\hfill\break
CDF obtained the masses of the $\Sigma_b^-$ and
$\Sigma_b^+$ from the decay \ $\Sigma_b \rightarrow \Lambda_b + \pi$ \ by
measuring the corresponding mass differences
\cite{CDF_Sigma_b,CDFsigmab} 
\bea
M(\Sigma_b^-) - M(\Lambda_b) = 195.5^{+1.0}_{{-}1.0}\,({\rm stat.}) \pm
0.1\, \hbox{(syst.) MeV}
\nonumber\\
\\
M(\Sigma_b^+) - M(\Lambda_b) = 188.0^{+2.0}_{{-}2.3}\,({\rm stat.}) \pm
0.1\, \hbox{(syst.) MeV}
\nonumber
\eea
with isospin-averaged mass difference 
$M(\Sigma_b) - M(\Lambda_b) = 192$ MeV.
\hfill\break

The final values for $\Sigma_b^-$ and $\Sigma_b^+$ are \cite{CDF_Sigma_b}
\bea
M(\Sigma_b^-) = 5816^{+1.0}_{{-}1.0}\,({\rm stat.}) \pm 1.7\, \hbox{(syst.) MeV}
\nonumber\\
\\
M(\Sigma_b^+) = 5808^{+2.0}_{{-}2.3}\,({\rm stat.}) \pm 1.7\, \hbox{(syst.) MeV}
\nonumber
\eea
with isospin-averaged mass $M(\Sigma_b) = 5812$ MeV.
\hfill\break

There is also the prediction for the spin splittings, good to 5\%
\beq{nurationum}
M(\Sigma_b^*)- M(\Sigma_b) = 
{{M(B^*)- M(B)}\over{M(K^*)-M(K)}}\cdot [M(\Sigma^*)-M(\Sigma)]=
22 \,{\rm MeV} 
\end{equation}
\hfill\break
to be compared with 21  MeV from the 
isospin-average of CDF measurements\cite{CDF_Sigma_b}
\hfill\break
\bea
M(\Sigma_b^{*-}) = 5837^{+2.1}_{{-}1.9}\,({\rm stat.}) \pm 1.7\, \hbox{(syst.) MeV}
\nonumber\\
\\
M(\Sigma_b^{*+}) = 5829^{+1.6}_{{-}1.8}\,({\rm stat.}) \pm 1.7\, \hbox{(syst.) MeV}
\nonumber
\eea

\vbox{
\strut
\hfill\break
\hfill\break
\centerline{
\includegraphics[width=30em,clip=true,angle=90]{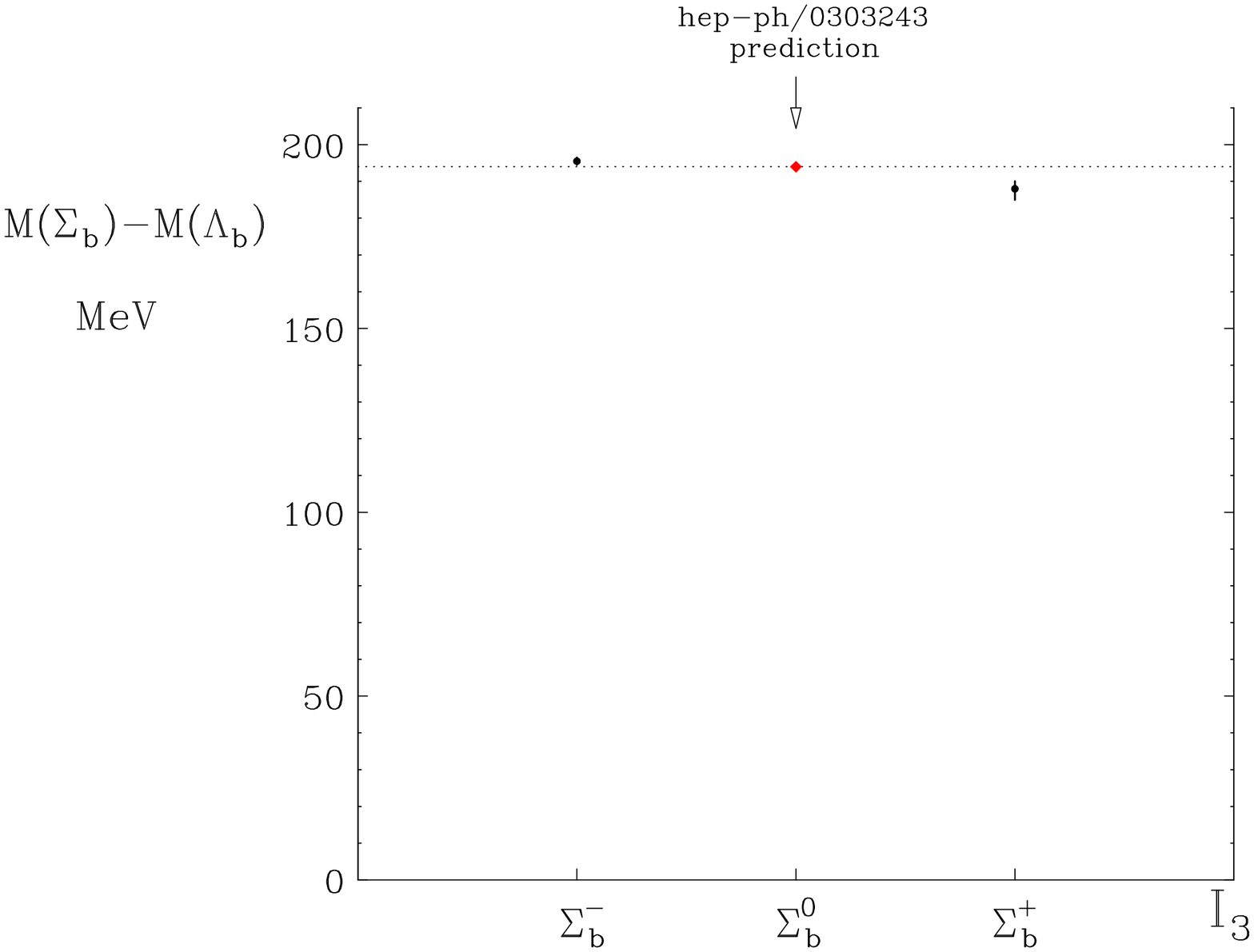}
}
\vskip0.5cm
{\small \em
Fig. 1. Experimental results from CDF for
$M(\Sigma_b^+)-M(\Lambda_b)$
and
$M(\Sigma_b^-)-M(\Lambda_b)$
compared with the theoretical prediction in Ref.~\cite{NewPenta}.
}
\hfill\break
} % end of \vbox

The success of the relation (\ref{newpred}) goes back to the question first raised by Andrei
Sakharov: {\em ``The $\Lambda$ and $\Sigma$ are made of the same quarks; 
why are their masses different?"}
One can now pose the same question for  all baryon
pairs denoted by $\Lambda_f$ and $\Sigma_f$, consisting of a $u$ and a $d$ quark
and a third quark $q_f$ of another flavor $f$ which can be $s$, $c$ or $b$. 
The spins are coupled differently and the hyperfine interaction between a $u$
quark and a $d$ quark is stronger than the hyperfine  interaction between one
$u$ or $d$  quark and $q_f$ quark. We also note that the hyperfine interaction
in mesons between a $u \bar d$ pair is stronger than the hyperfine interaction
between a $u$ or $d$ and an antiquark $\bar q_f$ of flavor $f$. 

The lhs of the relation (\ref{newpred}) takes the ratio of the $\Sigma_b-\Lambda_b$
mass difference which measures the hyperfine interaction difference between a 
$ud$ pair and a $ub$ or $db$ pair to the $\Sigma-\Lambda$
mass difference which measures the hyperfine interaction difference between a 
$ud$ pair and a $us$ or $ds$ pair. 
The rhs of the relation (\ref{newpred}) takes the ratio of
a combination of meson mass
differences which measures the hyperfine interaction difference $u \bar d$ pair
and a $u\bar b$ or $d\bar b$ pair. The relation (\ref{newpred}) is based on the
assumption that the $qq$ and $q \bar q$ interactions have the same flavor
dependence. This automatically follows from the assumption\cite{DGG} that both
hyperfine interactions are inversely proportional to the products of the same
quark masses. But all that is needed here is the weaker assumption of same
flavor dependence\cite{SakhZel},

\beq{qqbar}
{{V_{hyp}(q_i \bar q_j)}
\over{V_{hyp}(q_i \bar q_k)}}=
{{V_{hyp}(q_i q_j)}
\over{V_{hyp}(q_i q_k)}}
\end{equation}
The original derivation\cite{NewPenta} assumed that hyperfine interactions were
inversely proportional to the products of quark masses,
reflecting the fact that the spin-spin interaction is  linear in color-magnetic
moments of the quarks,  which in turn are inversely proportional to quark
masses,
\beq{SigLam4m}
\vrule height 9.5ex depth 1.5ex width 0pt
\kern-1ex
\left({\ds 1 - {m_u \over m_c}
\over
\vrule height 3.5ex depth 1.5ex width 0pt
\ds 1 - {m_u \over m_s}}\right)_{\strut \kern-1exBar}
\vrule height 1.5ex depth 2.5ex width 0pt 
\kern-3.0ex
= {{M_{\Sigma_c} - M_{\Lambda_c}}\over{M_{\Sigma} - M_\Lambda}}=2.16
=
\left({\ds 1 - {m_u \over m_c}
\over
\vrule height 3.5ex depth 1.5ex width 0pt
\ds 1 - {m_u \over m_s}}\right)_{\strut \kern-1exMes}
\vrule height 1.5ex depth 2.5ex width 0pt
\kern-3.0ex
={{(M_\rho {-} M_\pi){-}(M_{D^*}{-}M_D)}
\over
{(M_\rho {-} M_\pi){-}(M_{K^*}{-}M_K)}}
=2.10
\end{equation}

The simplicity of eq.(\ref{SigLam4m}) is somewhat misleading, because 
it hides the fact that the strength of the color hyperfine interaction also 
depends on the hadron wavefunction at the origin, which is
model-dependent\cite{boaz}. We avoid this difficulty by using the weaker
assumption eq. (\ref{qqbar}) which is model-independent and unlike quark
masses related to directly measurable observables.

Extending the relation (\ref{newpred}) to any two different 
flavors and rearranging the two sides to create baryon-meson ratios
gives\cite{NewPenta}
\begin{equation}
\begin{array}{ccccccc}
%\tallstrut
\displaystyle
{{M_{\Sigma_b} - M_{\Lambda_b}}\over{(M_\rho - M_\pi)-(M_{B^*}-M_B)}} &=&  
\displaystyle
{{M_{\Sigma_c}{-}M_{\Lambda_c}}\over{(M_\rho {-} M_\pi){-}(M_{D^*}{-}M_D)}}
&=&
\displaystyle
{{M_{\Sigma}{-}M_\Lambda}\over{(M_\rho {-} M_\pi){-}(M_{K^*}{-}M_K)}}
%\strut\kern-5em\strut
\label{eq:newpred2}
\\
\\
%[4pt]
%(5812-5624)/((775.8-139.57)- (5325.0 -5279.2))=
0.32
&\approx &
%(2452.03 -2286.46)/((775.8-139.57)-(2008.3-1866.9))=
0.33
&\approx& 
%(1193.15 - 1115.68)/((775.8-139.57)-(891.66-493.677))=
0.325
\end{array}
\end{equation}
The baryon-meson ratios are seen to be independent of the flavor $f$.

The challenge is to understand how and under what assumptions one can derive
from QCD  the very simple model of hadronic structure at  low energies which
leads to such accurate predictions. 

We shall present here many results relating meson and baryon masses  which have
been obtained without any  explicit model for the hyperfine interaction beyond
their flavor dependence. They relate experimental masses of mesons and baryons
containing quarks of   five different flavors $u,d,s,c,b$ with no free
parameters. It is difficult to believe  that these relations   are accidental
when they relate so many experimentally observed  masses of mesons and baryons.
This suggests that any model for hadron spectroscopy which treats mesons and
baryons differently or does not yield agreement with data for  all five flavors
is missing essential physics.   

That some kind of meson-baryon or light antiquark-diquark symmetry or
effective broken supersymmetry describes a number of relations between
meson and baryon masses has been noted\cite{szmassqcd}. The new successful
relations (\ref{eq:newpred2}) fit into this effective supersymmetry
picture. We now develop a formal description of this effective
supersymmetry \cite{szmassqcd} and obtain new relations between masses of
mesons and baryons. These relations do not have a simple description in
traditional QCD treatments which treat meson and baryon structures very
differently.

\def\MM{{\cal M}}
\def\BB{{\cal B}}

\section{The LS transformation -  A new meson-baryon supersymmetry?}
\subsection{The prediction for the newly discovered  $\Sigma_b$ baryons}

That meson and baryon masses must be related because they are made of the same
quarks was first pointed out by Sakharov and Zeldovich\cite{SakhZel} in a paper
that was completely ignored until the same work was independently
rediscovered\cite{hjl78}. 

We go beyond the simple quark model to find clues to the nonperturbative
dynamics of QCD. We search for the
minimum set of assumptions needed to derive old and new successful relations 
between mesons and baryons.
This supersymmetry transformation goes beyond the simple constituent quark
model. It assumes only a valence quark of flavor $i$ with a model independent 
structure bound to
``light quark brown muck color antitriplet" of model-independent structure 
carrying the quantum numbers of a light antiquark or a light diquark.
Since it assumes no model for the valence quark, nor the brown muck antitriplet 
coupled to the valence quark,
it holds also for the quark-parton model in 
which the valence is carried by a current quark and the rest of the hadron 
is a complicated mixture of quarks and antiquarks. 

This light quark supersymmetry transformation, denoted here by $T^S_{LS}$,
connects a meson denoted by $\vert \MM(\bar q Q_i) \rangle$ 
and a baryon denoted by 
$\vert \BB({[qq]\kern-0.5ex\phantom{\meddeepstrut}}_S Q_i) \rangle$
both containing the same valence quark
of some fixed flavor $Q_i$, $i=(u,s,c,b)$ and 
a light color-antitriplet ``brown muck" state with 
the flavor and baryon quantum numbers respectively of an antiquark $\bar 
q$ ($u$ or $d$) and two light quarks coupled to a diquark of spin $S$. 
\beq{numesbar}
T^S_{LS} \vert \MM(\bar q Q_i) \rangle
\quad \equiv \quad 
\vert \BB({[qq]\kern-0.5ex\phantom{\meddeepstrut}}_S Q_i) \rangle
\end{equation}
The mass difference between the meson and baryon related by this $T^S_{LS}$
transformation has been shown \cite{szmassqcd} to be independent of the quark
flavor $i$ for all four flavors $(u,s,c,b)$ when the contribution of the 
hyperfine interaction energies is removed.
For the two cases of spin-zero\cite{szmassqcd} $S=0$ and 
spin-one $S=1$ diquarks,   
\begin{equation}
\begin{array}{ccccccc}
%\tallstrut
M(N) - \tilde M(\rho) &=& M(\Lambda) - \tilde M(K^*) &=&
M(\Lambda_c) -  \tilde M(D^*) &= &
M(\Lambda_b) - \tilde M(B^*)
\\
%[4pt]
323~\rm{MeV} &\approx& 321~\rm{MeV} &\approx&312~\rm{MeV}
&\approx& 310~\rm{MeV}
\end{array}
\label{eq:mesbardif}
\end{equation}
\begin{equation}
\begin{array}{ccccccc}
%\tallstrut
\tilde M(\Delta) - \tilde M(\rho) &=& \tilde M(\Sigma) - \tilde M(K^*) &=&
\tilde M(\Sigma_c) -  \tilde M(D^*) &= &
\tilde M(\Sigma_b) - \tilde M(B^*)
\\
%[4pt]
517.56 ~\rm{MeV} &\approx& 526.43 ~\rm{MeV} &\approx& 523.95  ~\rm{MeV}
&\approx& 512.45 ~\rm{MeV}
\end{array}
\label{eq:mesbardif2}
\end{equation}
where 
\beq{tildev}
\tilde M(V_i)\equiv {{3M_{\MUU_i} + M_{\MUD_i} }\over {4}}; 
\end{equation}
are the weighted averages of vector and pseudoscalar meson masses,
denoted respectively by $M_{\MUU_i}$ and $M_{\MUD_i}$, which cancel their
hyperfine contribution,
and 
\beq{tildeb}
\tilde M(\Sigma_i)\equiv {{2M_{\Sigma^*_i} + M_{\Sigma_i}}\over {3}} ;
\qquad
\tilde M(\Delta)\equiv{{2 M_\Delta +M_N}\over {3}}
\end{equation} 
are the analogous weighted averages
of baryon masses which cancel the hyperfine contribution between the diquark
and the additional quark. 

We also note the striking constancy of the difference between  eqs.
(\ref{eq:mesbardif2}) and (\ref{eq:mesbardif}),
which gives the variation of
the spin splitting of the nonstrange diquark in baryons with different
companion  quarks, 

\begin{equation}
\begin{array}{ccccccc}
%\tallstrut
\tilde M(\Delta) -M(N)  &=& \tilde M(\Sigma) -M(\Lambda)  &=&
\tilde M(\Sigma_c) - M(\Lambda_c)&= &
\tilde M(\Sigma_b) -M(\Lambda_b) 
\\
%[4pt]
195~\rm{MeV} &\approx& 205 ~\rm{MeV} &\approx& 212 ~\rm{MeV}
&\approx& 202 ~\rm{MeV}
\end{array}
\label{eq:spindiq}
\end{equation}
\hfill\break

 The ratio of the hyperfine splittings of mesons and baryons 
related by $T^1_{LS}$ is also independent of the quark
flavor $i$ for all four flavors $(u,s,c,b)$,
\hfill\break

\beq{LShyperfine}
\begin{array}{ccccccc}
\displaystyle
\frac{M_\rho - M_\pi}{M_\Delta - M_N}&= &
\displaystyle
\frac{M_{K^*}-M_K}{M_{\Sigma^*} - M_\Sigma}&= &
\displaystyle
\frac{M_{D^*}-M_D}{M_{\Sigma_c^*} - M_{\Sigma_c}}&= &
\displaystyle
\frac{M_{B^*}-M_B}{M_{\Sigma^*_b}-M_{\Sigma_b}}
\\
2.17\pm0.01 &=& 2.08\pm0.01  &= & 2.18 \pm 0.01 &= & 2.15 \pm 0.20
\end{array}
\end{equation}
\hfill\break

That masses of boson and fermion states related by this transformation
(\ref{numesbar}) satisfy  simple relations like (\ref{eq:newpred2}), 
(\ref{eq:mesbardif}),
(\ref{eq:mesbardif2}) and
(\ref{LShyperfine})
remains a challenge for QCD, perhaps indicating some
boson-fermion or antiquark-diquark effective supersymmetry.

The meson and baryon ratios(\ref{SigLam4m})agree to $\pm 3\%$.
Eq. (\ref{SigLam4m}) is based on exactly the same logic as the prediction
(\ref{newpred}) which is accurate to $1\%$.  
The meson and baryon ratios
in eq. (\ref{LShyperfine}) differ over a range of 5\%.
These discrepancies at the level of several per cent presumably arise from
effects that are not included in our simple model.  Two such effects are:

\begin{enumerate} 
\item Neglect of electromagnetic contributions to hyperfine
interactions. These can produce small violations of the relation 
(\ref{qqbar}) for the case where the two quarks $q_j$ and $q_k$ have different
electric charges.
\item Neglect of differences between the wave functions of
spin-1/2 and  spin 3/2 baryons, 
\end{enumerate}

The  $\Sigma_b$ prediction (\ref{newpred}) is particularly
insensitive to these effects since it involves only spin-1/2 baryons and     
relates only hyperfine interactions of $b$ and $s$ quarks which have the same
electric charge and the same ratio of the strong to electromagnetic hyperfine
interactions.

The relations ~\eqref{eq:mesbardif} and ~\eqref{eq:mesbardif2} do not assume
any strengths for hyperfine interactions, only that their contributions are
canceled by suitable spin averaging. They are therefore also insensitive to the
electromagnetic contributions to the hyperfine interactions.

The relation (\ref{SigLam4m}) relates  hyperfine interactions of $c$ and $s$
quarks which have different electric charges and different ratios of the strong
to electromagnetic hyperfine interactions. This difference can easily account
for discrepancies of several per cent in experimental predictions

\subsection{Prediction for $\Xi_b$ and $\Xi_b^\prime$ baryons}

We can now extend this supersymmetry to apply to the case where the brown 
muck carries one unit of strangeness and has  
the flavor and baryon quantum numbers respectively of a 
strange antiquark $\bar 
s$ or a $us$ or $ds$ quark pair coupled to spin $S$.

\begin{equation}
\begin{array}{ccccccc}
%\tallstrut
%(1/4)M(\Lambda)+(3/4)M(\Sigma) - \tilde M(K^*) &= & 
M(\Xi_c) -  \tilde M(D_s^*) &= &
M(\Xi_b) - \tilde M(B_s^*)
\\
%(1/4)*(1116 + 3*1193) - 792.5 = 1174 -793 = 381 
%381 \rm{MeV} &\approx& 
394 ~\rm{MeV} &\approx& M(\Xi_b) - \tilde M(B_s^*)
\end{array}
\label{eq:mesxidif}
\end{equation}

\begin{equation}
\begin{array}{ccccccc}
%\tallstrut
%(2/3)M(\Sigma^*) + (1/3)[(1/4)M(\Sigma)+(3/4)M(\Lambda)]  - \tilde M(K^*) &= &
%(2/3)*1385 + (1/3)*(1/4)*(3*1116 + 1193) - 792.5 = 
% 923 + (1/4)*1514 -792.5 = 1301.5 - 792.5 =  509 
\tilde M(\Xi'_c) -  \tilde M(D_s^*) &= &
\tilde M(\Xi'_b) - \tilde M(B_s^*)
\\
% (2/3)*2643+(1/3)*2576.9 - 2076 = 2621-2076 =&\approx& 545 ~\rm{MeV} 
%509 &\approx& 
545 ~\rm{MeV} &\approx&
\tilde M(\Xi'_b) - \tilde M(B_s^*)
\end{array}
\label{eq:mesxiprdif}
\end{equation}

 These predict 
\begin{equation}
\begin{array}{ccccccc}
%\tallstrut
%\tilde
M(\Xi_b) \approx M(\Xi_c) -  \tilde M(D_s^*) + \tilde M(B_s^*)
= \tilde M(B_s^*) + 394~\rm{MeV} = 5795  ~\rm{MeV}
\\
\tilde M(\Xi'_b)\approx \tilde M(\Xi'_c) -  \tilde M(D_s^*) + \tilde M(B_s^*)
=  \tilde M(B_s^*) + 545~\rm{MeV}= 5950  ~\rm{MeV}
\end{array}
\label{eq:xipred}
\end{equation}
Closely related work focusing on estimating the 
$\Xi_b$ mass appeared recently in Ref. \cite{KKLR}, with the prediction 
$M(\Xi_b) = 5795\pm5$ MeV.
The recent CDF value $M(\Xi_b^-) =  5792.9 \pm 2.4 \pm 1.7  ~\rm{MeV}$ 
\cite{Litvintsev:2007,CDFxib},
announced after Ref.~\cite{KKLR} appeared,
is in surprising agreement with this prediction and with 
the result~\eqref{eq:xipred}. These results are also consistent with the 
D0 value $M(\Xi_b^-)= 5774   \pm 11   \pm 15$ MeV \cite{Abazov:2007ub}.

Both experiments are also consistent with a prediction in 
Ref.~\cite{Jenkins:1996de}, $M(\Xi_b) = M(\Lambda_b) + (182.7 \pm 5.0)$ MeV $ =
(5802.4 \pm 5.3)$ MeV.
But we have used both meson and baryon masses as input and only considered
hadrons containing strange quarks, while Ref.~\cite{Jenkins:1996de},
has used input only from baryon masses but also included nonstrange baryons.
Any differences in these predictions can pinpoint the relative importance of
deviations from assumed strangeness dependence and from assumed  meson-baryon
supersymmetry.

That the value of $m_b-m_c$ obtained from $B$ and $D$ mesons depends  upon the
flavor of the spectator quark was noted in Ref.~\cite{Karliner:2003sy}.  Table
I of Ref.~\cite{Karliner:2003sy} shows that the value of the effective quark
mass difference  $m_b-m_c$ obtained from experimental hadron masses  is the
same for mesons and baryons not containing strange quarks but different when
obtained from $B_s$ and $D_s$ mesons. Some reasons for this difference were
noted and the issue requires further investigation.

The values $394 ~\rm{MeV}$ and $545 ~\rm{MeV}$  in eqs.~\eqref{eq:mesxidif} and
\eqref{eq:mesxiprdif} can be considered as an  effective mass difference
between $(us)$ diquarks with respectively spin zero and spin one and a strange
antiquark. This gives $545 - 394 \approx 151~\rm{MeV}$ for the hyperfine
splitting of a $(us)$ diquark. They can be compared with the corresponding
values $312~\rm{MeV}$ and  $520~\rm{MeV}$ for the effective mass difference
between $(ud)$ diquarks with respectively spin zero and spin one and a
nonstrange antiquark. This gives $520 - 312 \approx 208~\rm{MeV}$ for the
hyperfine splitting of a $(ud)$ diquark and $(208/151) \approx 1.4$ for the
ratio of the two hyperfine splittings. This is in reasonable agreement with
values for the strangeness dependence of hyperfine splittings obtained from
other data. 

The hyperfine
interaction between the heavy quark, $c$ or $b$ acts differently on the spins
of the $u$ and $s$ quarks in the strange diquark. The small
mixing produced between the baryon states containing $(us)$ diquarks with spin 
zero and
spin  one\cite{Maltman:1980er,Lipkin:1981,Rosner:1981yh,Rosner:1992qa} 
has now been shown to be negligible\cite{KKLR}.

\subsection{Extending the supersymmetry to doubly strange diquarks}

We can now extend the effective supersymmetry to the case of
 hadrons related by changing a strange antiquark
$\bar s$ to a doubly strange $ss$ diquark coupled to spin $S = 1$.
\begin{equation}
\begin{array}{ccccccc}
\tallstrut
\displaystyle
\displaystyle
{{M(\Xi^*){-}M(\Xi)}\over{M(K^*){-}M(K)}}
&=&
\displaystyle
{{M(\Omega_c^*){-}M(\Omega_c)}\over{M(D_s^*){-}M(D_s)}}
%\strut\kern-5em\strut
\label{eq:newpredo}
\\
\\
0.54
&\approx& 
0.50 
\end{array}
\end{equation}

There is a remarkably successful relation for the spin-averaged doubly stranged
and charmed-doubly-strange baryons 

\begin{equation}
\begin{array}{ccccccc}
%\tallstrut
 \tilde M(\Xi) - \tilde M(K^*) &=&
\tilde M(\Omega_c) -  \tilde M(Ds^*) 
% &= & \tilde M(\Omega_b) - \tilde M(Bs^*)
\\
%[4pt]
%1461.3 - 792.5 = 668.8 & 2744.7 - 2075.5 = & X - 5405.
668.8 ~\rm{MeV} &\approx&  669.2~\rm{MeV}
% &\approx&  X - 5405 ~\rm{MeV}
\end{array}
\label{eq:mesbardifo}
\end{equation}
These relations have been recently extended to $b$-flavored hadrons\cite{KKLRb}.

\section{Magnetic moments of heavy quark baryons}

In $\Lambda$, $\Lambda_c$ and $\Lambda_b$ baryons the light quarks are
coupled to
spin zero. Therefore the magnetic moments of these baryons are determined
by the
magnetic moments of the $s$, $c$ and $b$ quarks, respectively. The latter
are
proportional to the chromomagnetic moments which determine the hyperfine
splitting
in baryon spectra. We can use this fact to
predict the $\Lambda_c$ and $\Lambda_b$ baryon magnetic moments by
relating them to the
hyperfine splittings in the same way as given in the original DGG
\cite{DGG}
prediction of the $\Lambda$ magnetic moment,

\beq{maglam}
\verytallstrut
\mu_\Lambda=
-{\mu_p\over 3}\cdot {{M_{\Sigma^*} - M_\Sigma} \over{M_\Delta - M_N}}
=-0.61 \,{\rm n.m.}
\qquad(\hbox{EXP}  =-0.61 \,{\rm n.m.})
\end{equation}

We obtain
\beq{maglamc}
\verytallstrut
\mu_{\Lambda_c}= -2 {\mu_\Lambda}\cdot
{{M_{\Sigma_c^*} - M_{\Sigma_c}}\over{M_{\Sigma^*} - M_\Sigma}}
=0.43\,{\rm n.m.}
\end{equation}
\beq{maglamb}
\verytallstrut
\mu_{\Lambda_b}= {\mu_\Lambda}\cdot
{{M_{\Sigma_b^*} - M_{\Sigma_b}} \over{M_{\Sigma^*} - M_{\Sigma}}}
=-0.067 \,{\rm n.m.}
\end{equation}

\section*{Acknowledgments}

We thank
Tom LeCompte,
Petar Maksimovic,
and
Vaia Papadimitriou
for correspondence on the experimental data and
Boaz Keren-Zur and Jon Rosner for useful discussions.

The research of one of us (M.K.) was supported in part by a grant from the
Israel Science Foundation administered by the Israel Academy of Sciences and
Humanities. The research of one of us (H.J.L.) was supported in part by the
U.S. Department of Energy, Division of Nuclear Physics, under contract number
DE-AC02-06CH11357.

%----------------------------------------------------------------------
% This prevents REFERENCES from forcing a page break
%\def\newpage{\vskip10ex}
%
\catcode`\@=11 % This allows us to modify PLAIN macros
\def\references{
\ifpreprintsty \vskip 10ex
%\ifpreprintsty \newpage
%
\hbox to\hsize{\hss \large \refname \hss }\else
\vskip 24pt \hrule width\hsize \relax \vskip 1.6cm \fi \list
{\@biblabel {\arabic {enumiv}}}
{\labelwidth \WidestRefLabelThusFar \labelsep 4pt \leftmargin \labelwidth
\advance \leftmargin \labelsep \ifdim \baselinestretch pt>1 pt
\parsep 4pt\relax \else \parsep 0pt\relax \fi \itemsep \parsep \usecounter
{enumiv}\let \p@enumiv \@empty \def \theenumiv {\arabic {enumiv}}}
\let \newblock \relax \sloppy
  \clubpenalty 4000\widowpenalty 4000 \sfcode `\.=1000\relax \ifpreprintsty
\else \small \fi}
\catcode`\@=12 % at signs are no longer letters
%-----------------------------------------------------------------
%{\tighten

  % end of global \tighten

\begin{thebibliography}{99}

\bibitem{NewPenta}
M. Karliner and H.J. Lipkin, arXiv:hep-ph/0307243; condensed version in 
Phys.\ Lett.\ B {\bf 575} (2003) 249.
\bibitem{CDF_Lambda_b}
D.~Acosta {\it et al.}  [CDF Collaboration],
  Phys.\ Rev.\ Lett.\  {\bf 96}, 202001 (2006),
  arXiv:hep-ex/0508022

\bibitem{CDF_Sigma_b}
{\tt http://www-cdf.fnal.gov/physics/new/bottom/060921.blessed-sigmab/}
\bibitem{CDFsigmab} T.~Aaltonen {\it et~al.} [CDF Collaboration],
  %``First Observation of Heavy Baryons \Sigma_b and \Sigma_b^*,''
  [arXiv:hep-ex/0706.3868].

\bibitem{RPP}
W.~M.~Yao {\it et al.}  [Particle Data Group],
  %``Review of particle physics,''
  J.\ Phys.\ G {\bf 33}, 1 (2006).

\bibitem{SakhZel}{
Ya.B. Zeldovich and A.D. Sakharov, Yad. Fiz {\bf 4}(1966)395; 
Sov. J. Nucl. Phys. {\bf 4}(1967)283.}

\bibitem {hjl78} Harry J. Lipkin, 
%Magnetic moments, hadron masses and quark masses. 
Phys. Rev. Lett. {\bf 41}, 1629 (1978).

\bibitem{szmassqcd}
M. Karliner and H.J. Lipkin, arXiv:hep-ph/0608004, Phys. Lett. B, in press.


\bibitem {hjlic} Harry J. Lipkin and I. Cohen. 
%Why masses and magnetic moments satisfy naive quark model predictions
Phys. Lett. {\bf 93B}, 56 (1980).


\bibitem{DGG}{A. De Rujula, H. Georgi and S.L. Glashow, Phys. Rev. D12
(1975) 147}

\bibitem {boaz} Boaz Keren-Zur, arXiv:hep-ph/070301, Ann. Phys., in press.
\bibitem{Maltman:1980er} K.~Maltman and N.~Isgur,
  Phys.\ Rev.\ D {\bf 22}, 1701 (1980).
\bibitem{Lipkin:1981} H. J. Lipkin, unpublished.
\bibitem{Rosner:1981yh} J.~L.~Rosner,
  Prog.\ Theor.\ Phys.\ {\bf 66}, 1422 (1981).
\bibitem{Rosner:1992qa} J.~L.~Rosner and M.~P.~Worah,
  Phys.\ Rev.\ D {\bf 46}, 1131 (1992).

\bibitem{KKLR}
Marek Karliner, Boaz Keren-Zur, Harry J. Lipkin
and Jonathan L. Rosner,  arXiv:hep-ph/0706.2163.

\bibitem{Litvintsev:2007} D. Litvintsev, on behalf of the CDF Collaboration,
seminar at Fermilab, June 15, 2007,
{\tt http://theory.fnal.gov/jetp/talks/litvintsev.pdf}\ .

\bibitem{Abazov:2007ub} V.~Abazov {\it et al.} [D0 Collaboration],
  %``Direct observation of the strange $b$ baryon $\Xi_b^{-}$,''
  [arXiv:hep-ex/0706.1690], Phys.\ Rev.\ Lett.\ {\bf 99}, 052001 (2007).

\bibitem{CDFxib} T.~Aaltonen {\it et~al.} [CDF Collaboration],
  %``Observation and Mass Measurement of the Baryon $\Xi^-_b$,''
  [arXiv:hep-ex/0707.0589], Phys.\ Rev.\ Lett.\ {\bf 99}, 052002 (2007).

\bibitem{Jenkins:1996de} E.~Jenkins,
  %``Heavy Baryon Masses in the $1/m_Q$ and $1/N_c$ Expansions,''
  %``Update of heavy baryon mass predictions,''
  Phys.\ Rev.\ D {\bf 54}, 4515 (1996); {\it ibid.} {\bf 55}, 10 (1997).

\bibitem{Karliner:2003sy} M.~Karliner and H.~J.~Lipkin,
  %``The constituent quark model revisited: Quark masses, new predictions for
  %hadron masses and K N pentaquark,''
  arXiv:hep-ph/0307243 (unpublished).
 
\bibitem{KKLRb}
Marek Karliner, Boaz Keren-Zur, Harry J. Lipkin
and Jonathan L. Rosner, arXiv:hep-ph/0708.4027

\end{thebibliography}
\end{document}